\documentclass[11pt]{article}
\usepackage{latexsym}
\usepackage{amsmath}
\usepackage{amssymb}
\usepackage{amsthm}
\usepackage{epsfig}
\usepackage{graphicx}
\usepackage{cancel}
\usepackage{mathtools}
\usepackage{enumitem}
\usepackage{easyReview}
\usepackage{mathabx}
\usepackage{hyperref}
\usepackage{caption}

\allowdisplaybreaks
\advance \topmargin by -\headheight
\advance \topmargin by -\headsep
\evensidemargin \oddsidemargin
\begin{document}

\title{An Analytical Framework for Frequency-Dependent Electromagnetic Power Absorption in Biological Tissues}
\author{Hongyun Wang 
\\
Department of Applied Mathematics \\
University of California, Santa Cruz, CA 95064, USA\\
\\
Shannon E. Foley \\
U.S. Department of Defense \\
Joint Intermediate Force Capabilities Office \\
Quantico, VA 22134, USA\\
\\
Hong Zhou\footnote{Corresponding author, hzhou@nps.edu} \\ Department of Applied Mathematics\\
Naval Postgraduate School, Monterey, CA 93943, USA
}

\maketitle
\begin{abstract}

As exposure to electromagnetic waves becomes increasingly widespread, it is important to quantify how incident fields couple into biological tissue and where absorbed energy is deposited. This work presents an analytical, physics-based framework derived from Maxwell’s equations to model the propagation of a normally incident electromagnetic plane wave within homogeneous, lossy dielectric biological tissues. Closed-form expressions for the electric and magnetic fields are derived, enabling the determination of frequency-dependent power reflectance and transmittance at the air–tissue interface, as well as the power absorption coefficient and penetration depth within the medium. Using complex relative permittivity data from the literature, we examine six tissue types across a broad frequency range 
(1~MHz–-100~GHz). The results demonstrate that higher water content significantly increases dielectric loss and reduces penetration depth. Conversely, low-water tissues (e.g., non-infiltrated fat) exhibit lower attenuation and deeper penetration. Frequency is shown to be a dominant driver of this behavior, with higher frequencies shifting the power budget from reflection-limited coupling toward highly superficial absorption. These findings provide a foundation basis for exposure assessments and the design of emerging electromagnetic technologies. 
\end{abstract}
\section{Introduction}
\label{intro_section}
Electromagnetic waves underpin a wide range of modern technologies, including communications and networking, medical and healthcare devices, sensing and imaging systems, transportation and automotive platforms, and defense and security applications  \cite{Zhadobov_2011, Colombi_2017, Owda_2020, Wang_2024, Gas_2026}. When electromagnetic radiation encounters the human body, a fraction of the incident power is reflected at the air–tissue interface, while the remainder is transmitted into the tissue and is attenuated through absorption. Quantifying this partition of power and the resulting spatial localization of absorbed power is essential to model the thermal effect and to connect electromagnetic design choices to biologically relevant metrics such as penetration depth and power absorption coefficient.

As electromagnetic technologies continue to evolve, understanding their biological effects and safety implications has become increasingly important. Millimeter waves (MMWs), typically in the range of 30 to 300 GHz, are of particular interest due to their expanding use in 5G and future communication networks, imaging, sensing, security screening, and military operations. A substantial body of research has examined the mechanisms by which MMW radiation interacts with biological tissues (e.g., \cite{Sacco_2021, Vilagosh_2020, Adekola_2025,  Alek_2000, Alek_2008, Gandhi_Riazi_1986}). Due to the short wavelengths of MMWs and the rapid power attenuation in tissue, penetration depths are generally shallow, with absorption primarily confined to superficial layers. This superficial absorption manifests itself mainly as thermal loading, producing localized temperature increases that can influence physiological and cellular processes. While existing exposure guidelines, such as those established by the International Commission on Non-Ionizing Radiation Protection (ICNIRP), are intended to prevent harmful thermal effects, ongoing studies continue to explore potential non-thermal responses and longer-term consequences. These investigations underscore the need for physically interpretable models that clarify how tissue properties and radiation frequency shape energy coupling and deposition.

To address this need, particularly in regimes where superficial absorption and heating dominate, this paper develops an analytical framework for electromagnetic power absorption in biological tissues. The approach is based on Maxwell’s equations for normal-incidence plane-wave propagation into a homogeneous lossy dielectric, allowing closed-form expressions for electric and magnetic fields, interface power reflectance and transmittance, attenuation (power absorption coefficient), and penetration depth. 

The remainder of this work is organized as follows. Section 2 derives the governing relations and closed-form field solutions for a plane wave in a homogeneous medium, along with expressions for the power flux (the Poynting vector) and related absorption metrics. 
Section 3 applies this framework using experimental data of frequency-dependent complex relative permittivity from Gabriel et al. \cite{Gabriel_1996b} to characterize the power absorption trends in six biological tissues: dry skin, wet skin, lens cortex, non-infiltrated fat, liver, and muscle. The analysis spans the frequency range from 1 MHz to 100 GHz, encompassing the MMW region where superficial absorption dominates.
Finally, Section 4 concludes with a summary of the main findings.
\section{Electromagnetic Propagation and Attenuation in Biological Tissue}
%
\subsection{Complex Relative Permittivity}
Biological tissues exhibit frequency-dependent dielectric properties, reflecting multiple polarizations spanning a wide temporal range. 
The complex relative permittivity $\varepsilon_r$ quantifies the electromagnetic response of a material
relative to that of free space, as a function of angular frequency $\omega$. In Gabriel et~al.\@ (1996) \cite{Gabriel_1996b}, 
the frequency-dependent relative permittivity, $\varepsilon_r(\omega)$,  
is phenomenologically described by 
\begin{equation}
\varepsilon_r(\omega) = \varepsilon_\infty + \sum  
\frac{\Delta \varepsilon_n}{1+(i \omega \tau_n)^{(1-\alpha_n)}} 
+ \frac{\sigma_i}{i \omega \varepsilon_0} 
\label{rela_perm_G}
\end{equation}
which captures both the polarizations of various time scales and the ionic conduction. Here,
\begin{itemize}
\item $\varepsilon_\infty $ is the high-frequency limit of the relative permittivity; 
\item $\sigma_i$ is the static ionic conductivity 
(with units of $\text{C}^2 \text{kg}^{-1} \text{m}^{-3} \text{s}$ or $\text{S}/\text{m}$); 
\item $\varepsilon_0 = 8.8541878 {\times} 10^{-12}\, 
\text{C}^2 \text{kg}^{-1} \text{m}^{-3} \text{s}^2$ is the absolute permittivity of free space;  
\item $(\Delta \varepsilon_n, \tau_n, \alpha_n)$ are parameters that model the polarization
of the material in response to electromagnetic oscillations. 
The polarization process may involve multiple clusters of relaxation time scales, 
as represented by the summation 
in \eqref{rela_perm_G}. 
\end{itemize}
The values of these parameters for biological tissues are sourced from Gabriel et~al.\@ (1996) \cite{Gabriel_1996b}, who reported extensive experimental measurements of the frequency-dependent dielectric properties for a wide range of biological materials. These datasets serve as a standard reference for modeling the electromagnetic response of biological tissues in bioelectromagnetics and related applications. The specific tissue properties utilized in this study are detailed in Section 3. 
%

%
%
%
%

\subsection{Plane Wave Solution in a Homogeneous Medium}
\begin{figure}[!h]
\vskip 0.0cm
\centering
\includegraphics[width=3.5in]{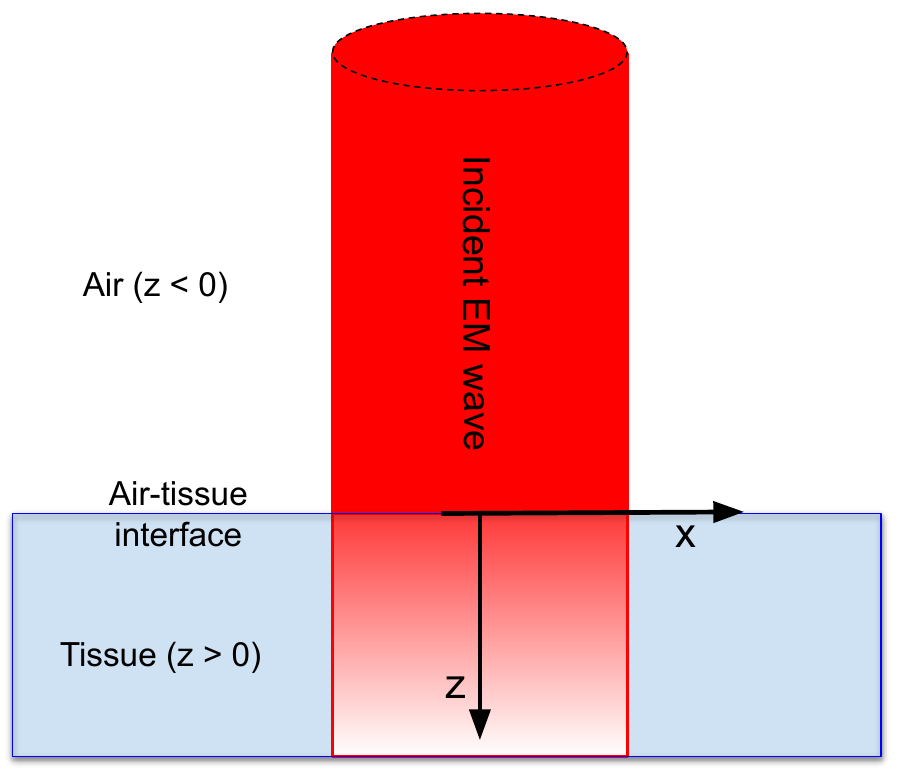}
\vskip -0.0cm
\caption{The coordinate system for an electromagnetic wave normally incident on a flat layer of biological tissue. $z$: propagation; 
$x$: electric field oscillation.}
\label{fig_0}
\end{figure}

Consider the case of a plane wave incident normally on a flat layer of biological tissue as shown in Figure \ref{fig_0}.
The two media (air and tissue) are separated by an infinite plane interface. 
The air is modeled as the lossless free space. 
Our goal is to analyze the reflection and transmission of the incident 
electromagnetic wave at the interface and the subsequent propagation and absorption inside the biological tissue, which results in thermal heating. 

To study the effect of interface, we first consider a plane wave propagating 
in a homogeneous medium filling the entire 3D space. 
We take the wave propagation direction as the $\mathbf{e}_z$, with the electric field
$\mathbf{E}$ oscillating along $\mathbf{e}_x$ and the magnetic field 
$\mathbf{H}$ along $\mathbf{e}_y$.
Let $\omega$ be the angular frequency of the plane wave.
With these setups, the time-harmonic electric and magnetic fields 
can be expressed as
\begin{equation}
\begin{dcases}
\mathbf{E}(\mathbf{x}, t) = e^{i\omega t } \mathbf{E}(z) = 
e^{i\omega t } E_1(z) \mathbf{e}_x \\ 
\mathbf{H}(\mathbf{x}, t) = e^{i\omega t } \mathbf{H}(z) = 
e^{i\omega t } H_2(z) \mathbf{e}_y 
\end{dcases}
\label{phasor_def}
\end{equation}
where $\mathbf{x} \equiv x \mathbf{e}_x + y \mathbf{e}_y + z \mathbf{e}_z$
denotes the 3D position vector.
Here, $\mathbf{E}(z)$ and $\mathbf{H}(z)$ represent the phasors
(also referred to as complex amplitudes) of the time-harmonic 
electric and magnetic fields, respectively.
The power flux, defined as the power per unit area transmitted through a wavefront,  
is characterized by the time-averaged Poynting vector. For time-harmonic electromagnetic fields, the instantaneous Poynting vector is given by the cross product of the electric and magnetic fields, and its physically relevant value is obtained by averaging over one temporal period. Accordingly, the time-averaged Poynting vector is
\begin{align*}
& \langle \mathbf{S} \rangle(\mathbf{x}) = \frac{1}{T_p} \int_0^{T_p} 
\text{Re}(\mathbf{E}(\mathbf{x}, t)) {\times} \text{Re}(\mathbf{H(\mathbf{x}, t)}) dt \\
& \qquad = \frac{1}{T_p} \int_0^{T_p} \frac{1}{2} \text{Re}\big(\mathbf{E}(\mathbf{x}, t) 
{\times} \mathbf{H}(\mathbf{x}, t)^*\big) dt \\
& \qquad = \frac{1}{2} \text{Re}\big(E_1(z) H_2(z)^*\big) \mathbf{e}_z 
\equiv \langle S_3 \rangle (z) \mathbf{e}_z 
\end{align*}
where $T_p \equiv \frac{2\pi}{\omega}$ is the temporal period of the fields. 
The time-averaged Poynting vector aligns with the direction of propagation, 
parallel to $\mathbf{e}_z$. Its $z$-component, $\langle S_3 \rangle(z)$, is positive or negative
depending on whether the wave propagates in the $+\mathbf{e}_z$ or $-\mathbf{e}_z$ direction, respectively. 
The following equations summarize the expressions for the time-averaged Poynting vector:
\begin{equation}
\langle \mathbf{S} \rangle(\mathbf{x}) = \langle S_3 \rangle(z) \mathbf{e}_z, \qquad 
\langle S_3 \rangle(z) \equiv \frac{1}{2} \text{Re}\big(E_1(z) H_2(z)^*\big) 
\label{S_exp_1} 
\end{equation}
To describe the behavior of electromagnetic fields at frequency $\omega$ within a linear, 
isotropic, and non-magnetic medium, we employ
Maxwell's equations in phasor form for $\mathbf{E}(z)$ and $\mathbf{H}(z)$:
\begin{align}
& \begin{dcases}
\nabla {\times} \mathbf{E}(z) =-i \omega \mu_0 \mathbf{H}(z) \\
\nabla {\times} \mathbf{H}(z) =
 i \omega \varepsilon_0 \varepsilon_r(\omega) \mathbf{E}(z) 
 \label{Mw_Eq_0}
\end{dcases}  
\end{align}
where $\mu_0 = 4\pi {\times} 10^{-7}\, \text{kg}\, \text{m}\, \text{C}^{-2} (\text{or }\text{H}\,/ \text{m})$ 
is the permeability of free space. Since biological tissues are non-magnetic, 
approximately we have $\mu = \mu_0$. 
In \eqref{Mw_Eq_0}, $\varepsilon_0 \varepsilon_r$ is the complex absolute permittivity of the tissue. Combining the expressions in
\eqref{Mw_Eq_0} yields the Helmholtz wave equation for the electric field phasor: 
\begin{align}
 \quad \nabla^2 \mathbf{E}(z) = -\omega^2 \mu_0 \varepsilon_0 
\varepsilon_r(\omega) \mathbf{E}(z) \label{Mw_Eq_1}
\end{align}
The propagation speed $c_0$ and the angular wavenumber $k_0$ in free space are defined as
\[ c_0 \equiv \frac{1}{\sqrt{\mu_0 \varepsilon_0}}, \qquad k_0 \equiv \frac{\omega}{c_0}
= \omega \sqrt{\mu_0 \varepsilon_0} \]
We can express the coefficients in \eqref{Mw_Eq_1} and \eqref{Mw_Eq_0}
in terms of the free-space wavenumber $k_0$:
\[ \omega^2 \mu_0 \varepsilon_0 = k_0^2, \qquad 
\omega \mu_0 = k_0 \eta_0, \quad \eta_0 \equiv \sqrt{\frac{\mu_0}{\varepsilon_0}} \] 
where $\eta_0 \equiv \sqrt{\frac{\mu_0}{\varepsilon_0}}$ is the intrinsic impedance of free space. 
For conciseness, we denote $\varepsilon_r(\omega)$ simply by $\varepsilon_r$. With these definitions, equations \eqref{Mw_Eq_0} and \eqref{Mw_Eq_1} can be rewritten as 
\begin{equation}
\begin{dcases}
\nabla^2 E_1(z) = -k_0^2 \varepsilon_r E_1(z) \\
 \partial_z(E_1(z)) = -i k_0 \eta_0 H_2(z)
\end{dcases} \label{Mw_Eq_2}
\end{equation}
The solution to \eqref{Mw_Eq_2} for a wave propagating in the positive $\mathbf{e}_z$ direction is given by
\begin{equation}
\begin{dcases}
E_1(z) = E_1^{(0)} e^{-\gamma z}, \qquad \gamma = i k_0 \sqrt{\varepsilon_r} \\
H_2(z) = \frac{1}{\eta_0} \sqrt{\varepsilon_r} E_1^{(0)} e^{-\gamma z}
\end{dcases} \label{Mw_sol_2}
\end{equation}
where $\gamma$ is the complex propagation constant. By convention, 
$\sqrt{\varepsilon_r}$ denotes the principal branch of the square root, ensuring a nonnegative real part for the refractive index and a nonnegative attenuation constant. 
The $z$-component of the time-averaged Poynting vector is 
\begin{align}
& \langle S_3 \rangle(z) = \frac{1}{2} \text{Re}(E_1(z) H_2(z)^*)  
= \frac{1}{2\eta_0} \text{Re}\big(\sqrt{\varepsilon_r}\big) 
\big| E_1^{(0)} \big|^2 e^{-2\text{Re}(\gamma) z} \label{S_exp_2}
\end{align}
The solution to \eqref{Mw_Eq_2} for a wave propagating in the negative $\mathbf{e}_z$ direction is  
\begin{equation}
\begin{dcases}
E_1(z) = E_1^{(0)} e^{+\gamma z}, \qquad \gamma = i k_0 \sqrt{\varepsilon_r} \\
H_2(z) = -\frac{1}{\eta_0} \sqrt{\varepsilon_r} E_1^{(0)} e^{+\gamma z}
\end{dcases} \label{Mw_sol_2B}
\end{equation}
This solution is employed when analyzing reflection at an interface. Notably, \eqref{Mw_sol_2B} differs from the forward-propagating case in \eqref{Mw_sol_2} by the reversal of the exponential argument ($e^{+\gamma z}$ instead of $e^{-\gamma z}$) and the sign change in the magnetic field component, reflecting the directional reversal of the power flux.

\subsubsection{Plane Wave Solution in Free Space}
In free space, where $\varepsilon_r = 1$ and $\gamma = i k_0$, 
a plane wave propagating in the $+\mathbf{e}_z$ direction is 
\begin{equation}
\begin{dcases}
E_1^{(I)}(z) = E_1^{(0,I)} e^{-i k_0 z} \\
H_2^{(I)}(z) = \frac{1}{\eta_0} E_1^{(0,I)} e^{-i k_0 z}, \qquad 
\eta_0 \equiv \sqrt{\frac{\mu_0}{\varepsilon_0}} \\
\langle S_3^{(I)} \rangle (z) =
\frac{1}{2 \eta_0} \big| E_1^{(0,I)} \big|^2 
\end{dcases} \label{Mw_sol_3I}
\end{equation}
For normal incidence on a biological tissue, 
\eqref{Mw_sol_3I} represents the incident wave. 
The superscript $^{(I)}$ is used to distinguish the incident field from the reflected and transmitted components.

Similarly, a wave propagating in the $(-\mathbf{e}_z)$ direction is given by
\begin{equation}
\begin{dcases}
E_1^{(R)}(z) = E_1^{(0,R)} e^{i k_0 z} \\
H_2^{(R)}(z) = -\frac{1}{\eta_0} E_1^{(0,R)} e^{i k_0 z}, \qquad 
\eta_0 \equiv \sqrt{\frac{\mu_0}{\varepsilon_0}} \\
 \langle S_3^{(R)} \rangle (z) = -
\frac{1}{2 \eta_0} \big| E_1^{(0,R)} \big|^2 
\end{dcases} \label{Mw_sol_3R}
\end{equation}
Equation \eqref{Mw_sol_3R} corresponds to the reflected wave, indicated by the superscript $^{(R)}$. The negative sign in $\langle S_3^{(R)} \rangle (z)$ 
reflects that power flows in the negative $\mathbf{e}_z$ direction. 

\subsubsection{Plane Wave Solution in a Lossy Medium}
In a lossy medium, the complex relative permittivity is expressed as $\varepsilon_r = \varepsilon' 
- i \varepsilon''$, where $\varepsilon'' > 0$ represents the dielectric loss. 
Its square root can be written as
\[ \sqrt{\varepsilon_r} = n - i \kappa \]
where $n$ is the refractive index that determines the phase velocity of the wave, and 
$\kappa$ is the extinction coefficient that governs the spatial attenuation of the wave. The condition
$\varepsilon'' > 0$ ensures that $\kappa> 0$, consistent with a passive, energy-dissipating medium. 
The complex propagation constant $\gamma $ is
\[ \gamma = i k_0 \sqrt{\varepsilon_r} = k_0 \kappa + i k_0 n \qquad \] 
Based on these definitions, a plane wave propagating in the $+\mathbf{e}_z$ direction is 
\begin{equation}
\begin{dcases}
E_1^{(T)}(z) = \big(E_1^{(0,T)}e^{-k_0 \kappa z}\big) e^{-i k_0 n z}, \qquad
(n - i \kappa) \equiv  \sqrt{\varepsilon_r} \\
H_2^{(T)}(z) = \big(\frac{ \sqrt{\varepsilon_r}}{\eta_0} 
E_1^{(0,T)}e^{-k_0 \kappa z}\big) e^{-i k_0 n z}, \qquad 
\eta_0 \equiv \sqrt{\frac{\mu_0}{\varepsilon_0}} \\
\langle S_3^{(T)} \rangle (z) =  \frac{1}{2 \eta_0} n \big| E_1^{(0,T)} \big|^2 e^{-2k_0 \kappa z}
\end{dcases} \label{Mw_sol_3T}
\end{equation}
Equation \eqref{Mw_sol_3T} corresponds to the wave that is transmitted into and continues propagating through the lossy medium; as such, it is labeled with the superscript $^{(T)}$.

\subsection{Reflection and Transmission of a Normally Incident Wave 
on a Flat Biological Tissue}
Consider a plane wave incident normally on a flat layer of biological tissue. 
At the air-tissue interface, the incident wave is partitioned into a reflected wave 
and a transmitted wave. 

We define the unit normal to the interface, pointing into the tissue, as the 
positive $\mathbf{e}_z$ direction. 
The incident wave propagates toward the positive $\mathbf{e}_z$ axis, with its
electric field polarized along $\mathbf{e}_x$ and its
magnetic field along $\mathbf{e}_y$. 
Within this coordinate system, the incident, reflected, and transmitted waves are characterized by equations \eqref{Mw_sol_3I}, \eqref{Mw_sol_3R}, and  \eqref{Mw_sol_3T}, respectively.

At the air-tissue interface, the infinitesimally thin boundary surface does not support an ionic current.
As a result, the tangential components of both the electric and magnetic fields must be continuous. This requirement yields the following boundary conditions: 
\begin{align*}
& \begin{dcases}
E_1^{(I)}(0) + E_1^{(R)}(0) = E_1^{(T)}(0) \\
H_2^{(I)}(0) + H_2^{(R)}(0) = H_2^{(T)}(0)
\end{dcases} \quad \Longrightarrow \quad 
\begin{dcases}
E_1^{(0,I)} + E_1^{(0,R)} = E_1^{(0,T)} \\
E_1^{(0,I)}- E_1^{(0,R)} = \sqrt{\varepsilon_r} E_1^{(0,T)} \\[1ex]
\end{dcases} 
\end{align*}
Solving this $2\times 2$ linear system for $E_1^{(0,T)}$ and $E_1^{(0,R)}$ in terms of $E_1^{(0,I)}$, we obtain 
\begin{equation}
\boxed{\quad \begin{dcases}
E_1^{(0,T)} = a_T E_1^{(0,I)}, \qquad \underbrace{a_T \equiv 
\frac{E_1^{(0,T)}}{E_1^{(0,I)}} = 
\frac{2}{1+\sqrt{\varepsilon_r}}}_{\text{field transmission coefficient}} \\[1ex]
E_1^{(0,R)} = a_R E_1^{(0,I)}, \qquad \underbrace{a_R \equiv 
\frac{E_1^{(0,R)}}{E_1^{(0,I)}} = 
\frac{1-\sqrt{\varepsilon_r}}{1+\sqrt{\varepsilon_r}}
}_{\text{field reflection coefficient}}
\end{dcases} \quad}
\label{aT_aR_eq}
\end{equation}
Note that, in general, both the field transmission coefficient $a_T$ and 
the field reflection coefficient $a_R$ are complex-valued. 
At the air-tissue interface ($z=0$), the $z$-components of the time-averaged Poynting vectors for the incident, reflected, and transmitted waves are given by
\[ \begin{dcases} 
 \langle S_3^{(I)} \rangle (0) = \frac{1}{2\eta_0} \big| E_1^{(0,I)} \big|^2 \\
 \big| \langle S_3^{(R)} \rangle (0) \big| = \frac{1}{2\eta_0} \big| a_R \big|^2 \big| E_1^{(0,I)} \big|^2 \\
\langle S_3^{(T)} \rangle (0) =  \frac{1}{2\eta_0} n \big| a_T \big|^2 \big| E_1^{(0,I)} \big|^2 
\end{dcases} \] 
%
The power transmission coefficient (transmittance) $\mathcal{T}$ and the reflectance $\mathcal{R}$ are defined as the ratios of the $z$-components of the time-averaged Poynting vectors at the interface. These power coefficients are expressed as
\begin{equation}
\boxed{\quad 
\mathcal{T} \equiv \frac{\langle S_3^{(T)} \rangle (0)}{\langle S_3^{(I)} \rangle (0)}
= n \big| a_T \big|^2 
= \text{Re}\big(\sqrt{\varepsilon_r} \big) \Big| \frac{2}{1+\sqrt{\varepsilon_r}} \Big|^2 
\quad }
\label{pow_transmission}
\end{equation}
\begin{equation}
\boxed{\quad 
\mathcal{R} \equiv \frac{\big| \langle S_3^{(R)} \rangle (0) \big|}{\langle S_3^{(I)} \rangle (0)}
= \big| a_R \big|^2 
= \Big| \frac{1-\sqrt{\varepsilon_r}}{1+\sqrt{\varepsilon_r}} \Big|^2 
\quad }
\label{reflectance}
\end{equation}
In the definition of reflectance $\mathcal{R}$, the absolute value is taken because
the $z$-component of the reflected Poynting vector, $\langle S_3^{(R)} \rangle (0)$, is negative; this 
indicates that power in the reflected wave flows in the $(-\mathbf{e}_z)$ direction. 
It is straightforward to verify that $ \mathcal{T} + \mathcal{R} = 1$, a result consistent with energy conservation at the air-tissue interface. 

Inside the biological tissue, the time-averaged Poynting vector has the expression 
\begin{equation}
\langle \mathbf{S}^{(T)} \rangle (z) =  \langle S_3^{(T)} \rangle (0)
e^{-2k_0 \text{Im}(-\sqrt{\varepsilon_r}) z} \mathbf{e}_z 
\label{S_exp_3} 
\end{equation}
where $\langle S_3^{(T)} \rangle (0)$ is the power flux evaluated at the interface.
In a lossy medium, $\varepsilon_r = \varepsilon' -i \varepsilon''$ with $\varepsilon'' > 0$, 
and thus $\text{Im}(-\sqrt{\varepsilon_r}) > 0$. 
It follows that  the exponential term in \eqref{S_exp_3} 
describes the spatial attenuation of power as the wave propagates deeper into the tissue.

In the electromagnetic heating of biological tissues, the attenuation of the power flux as it propagates through the tissue corresponds to the conversion of electromagnetic energy into heat. The spatial attenuation rate of the time-averaged Poynting vector $\langle \mathbf{S}^{(T)} \rangle (z)$ with respect to depth $z$ thus defines
the power absorption coefficient $\mu_\text{absorp}$. 
\begin{equation}
\boxed{\quad \langle \mathbf{S}^{(T)} \rangle (z) \; \propto \; e^{-\mu_\text{absorp}z}, 
\qquad 
\mu_\text{absorp}=\frac{2\omega}{c_0} \text{Im}(-\sqrt{\varepsilon_r}) \quad }
\label{aborp_coef}
\end{equation}
Note that the results derived above can be expressed in terms of the intrinsic impedances of air (free space) 
and biological tissue. Given that the absolute permittivity of the tissue is $\varepsilon_0 \varepsilon_r$, the intrinsic impedance of the medium is
\[ \eta = \sqrt{\frac{\mu_0}{\varepsilon_0 \varepsilon_r}} = \frac{\eta_0}{\sqrt{\varepsilon_r}} \]
which immediately gives 
\[
 \sqrt{\varepsilon_r} = 
\frac{\eta_0}{\eta} \]
Equations \eqref{aT_aR_eq}, \eqref{reflectance}, and \eqref{aborp_coef} provide the 
expressions for the field transmission and reflection coefficients, as well as  
the power absorption coefficient and reflectance. These parameters can be concisely expressed in terms of the intrinsic impedances of air and biological tissue as follows:
\[ a_T \equiv \frac{E_1^{(0,T)}}{E_1^{(0,I)}} = \frac{2\eta }{\eta+\eta_0}, \qquad
a_R \equiv \frac{E_1^{(0,R)}}{E_1^{(0,I)}} = \frac{\eta-\eta_0}{\eta+\eta_0} \] 
\[ \mathcal{R} \equiv \frac{\big| \langle S_3^{(R)} \rangle (0) \big|}{\langle S_3^{(I)} \rangle (0)}
= \Big| \frac{\eta-\eta_0}{\eta+\eta_0} \Big|^2, \qquad 
\mu_\text{absorp}=\frac{2\omega}{c_0} \text{Im}(-\sqrt{\varepsilon_r}) \]
These expressions explicitly demonstrate that reflection and transmission at the interface are determined solely by the impedance mismatch between air and biological tissue. Furthermore, the attenuation and absorption of electromagnetic power within the medium
are determined by the complex nature of the tissue's intrinsic impedance. 
These formulas for the transmission and reflection coefficients are the well-known Fresnel equations \cite{Griffiths_book, Yeh_book, Jackson_book}. While these results are well established, our derivations provide a mathematically transparent and physically intuitive framework for analyzing energy deposition in biological tissues. 

\section{Dielectric Behaviors of Biological Tissues for Electromagnetic Waves from 1 MHz to 100 GHz }
We examine the dielectric response of six biological tissues to incident electromagnetic waves across a frequency range of 1 MHz to 100 GHz. The tissues analyzed include (i) dry skin, (ii) wet skin, 
(iii) lens cortex, (iv) non-infiltrated fat, (v) liver, and (vi) muscle. 
The frequency-dependent dielectric properties are parameterized using the four-pole Cole--Cole model at 37$^\circ$C, as reported by Gabriel {\it et al.} \cite{Gabriel_1996b}.
For each tissue, the complex relative permittivity $\varepsilon_r(\omega)$ is defined as the sum of a high-frequency permittivity limit $\varepsilon_\infty$, four dispersive relaxation terms characterized by $(\Delta\varepsilon_n,\tau_n,\alpha_n)$, and an ionic conductivity $\sigma$, as given in 
\eqref{rela_perm_G}. 

For the reader's convenience, Table~\ref{tab_1} lists the parameters from \cite{Gabriel_1996b} 
used to analyze the six biological tissues. First, the relative permittivity 
 $\varepsilon_r(\omega)$ is calculated via \eqref{rela_perm_G}
over the frequency range of 1 MHz to 100 GHz. 
Subsequently, at each frequency, the relative permittivity $\varepsilon_r(\omega)$ is used 
to evaluate the impedance mismatch at the air-tissue interface, as well as the attenuation 
and absorption within the tissue.
We follow the unit convention of Gabriel \emph{et al.} \cite{Gabriel_1996b}, where $\tau_1$ is reported in ps, $\tau_2$ in ns, $\tau_3$ in $\mu$s, and $\tau_4$ in ms.
\begin{table*}[h]
\captionsetup{labelfont=bf} 
\centering
\caption{Parameters for the six biological tissues studied in this work (from Gabriel \emph{et al.} \cite{Gabriel_1996b}). 
These parameters are used in the Cole--Cole dielectric model to calculate the relative permittivity
as a function of frequency, as described in \eqref{rela_perm_G}.}
\label{tab_1}
\vskip 6pt
\renewcommand{\arraystretch}{1.1}
\small
\begin{tabular}{|l|c|c|c|c|c|c|}
\hline \hline
Parameter & Dry Skin & Wet Skin & Lens cortex & Fat (non-infiltrated) & Liver & Muscle \\
\hline \hline
$\varepsilon_\infty$ & 4.0 & 4.0 & 4.0 & 2.5 & 4.0 & 4.0 \\ \hline
$\sigma$ (S/m) & $2.0\times10^{-4}$ & $4.0\times10^{-4}$ & $3.0\times10^{-1}$ & $1.0\times10^{-2}$ & $2.0\times10^{-2}$ & $2.0\times10^{-1}$ \\ 
\hline \hline
$\Delta\varepsilon_1$ & 32.0 & 39.0 & 42.0 & 3.0 & 39.0 & 50.0 \\ \hline
$\tau_1$ (ps) & 7.23 & 7.96 & 7.96 & 7.96 & 8.84 & 7.23 \\ \hline
$\alpha_1$ & 0.00 & 0.10 & 0.10 & 0.20 & 0.10 & 0.10 \\ 
\hline \hline
$\Delta\varepsilon_2$ & 1100 & 280 & 1500 & 15 & 6000 & 7000 \\ \hline
$\tau_2$ (ns) & 32.48 & 79.58 & 79.58 & 15.92 & 530.52 & 353.68 \\ \hline
$\alpha_2$ & 0.20 & 0.00 & 0.10 & 0.10 & 0.20 & 0.10 \\ 
\hline \hline
$\Delta\varepsilon_3$ & 0 & $3.0\times10^{4}$ & $2.0\times10^{5}$ & $3.3\times10^{4}$ & $5.0\times10^{4}$ & $1.2\times10^{6}$ \\ \hline
$\tau_3$ ($\mu$s) & -- & 1.59 & 159.15 & 159.15 & 22.74 & 318.31 \\ \hline
$\alpha_3$ & -- & 0.16 & 0.10 & 0.05 & 0.20 & 0.10 \\ 
\hline \hline
$\Delta\varepsilon_4$ & 0 & $3.0\times10^{4}$ & $4.0\times10^{7}$ & $1.0\times10^{7}$ & $3.0\times10^{7}$ & $2.5\times10^{7}$ \\ \hline
$\tau_4$ (ms) & -- & 1.592 & 15.915 & 7.958 & 15.915 & 2.274 \\ \hline
$\alpha_4$ & -- & 0.20 & 0.00 & 0.01 & 0.05 & 0.00 \\
\hline \hline
\end{tabular}
\end{table*}

Several dielectric quantities are plotted as functions of frequency for each tissue. 
To facilitate direct comparison, the vertical scale for each property is kept fixed across all six tissues. 
We focus on the following dielectric properties:
\begin{itemize}
\item Real part of the relative permittivity (dielectric constant): 
$\text{Re}(\varepsilon_r)$
\item Negative imaginary part of the relative permittivity 
(dielectric loss): $\text{Im}(-\varepsilon_r)$ 
\item Reflectance: $\mathcal{R} \equiv \frac{\big| \langle S_3^{(R)} 
\rangle (0) \big|}{\langle S_3^{(I)} \rangle (0)}
= \Big| \frac{1-\sqrt{\varepsilon_r}}{1+\sqrt{\varepsilon_r}} \Big|^2 $ 
\item Power transmission coefficient: $\mathcal{T} \equiv \frac{ \langle S_3^{(T)} 
\rangle (0) }{\langle S_3^{(I)} \rangle (0)}= 1- \mathcal{R}$
\item Power absorption coefficient inside the tissue: 
$\displaystyle \mu_\text{absorp}=\frac{2\omega}{c_0} \text{Im}(-\sqrt{\varepsilon_r})$
\item Electromagnetic penetration depth into the tissue: 
$1/\mu_\text{absorp}$
\end{itemize}
These quantities characterize the interaction between electromagnetic waves and biological tissues. The real part of the permittivity, $\text{Re}(\varepsilon_r)$, determines the tissue’s ability to store energy during electromagnetic oscillation, while the negative imaginary part, $\text{Im}(-\varepsilon_r)$, quantifies dielectric losses that convert electromagnetic power into heat. The reflectance, $\mathcal{R}$, represents the fraction of incident power reflected at the tissue interface, whereas the transmittance (the power transmission coefficient), $\mathcal{T}$, measures the fraction of power transmitted into the tissue. The power absorption coefficient, $\mu_\text{absorp}$, governs the rate at which transmitted power attenuates with depth, and its reciprocal defines the penetration depth, indicating how far electromagnetic power propagates before being attenuated by a factor of $e \approx 2.71828$. Together, these properties provide a comprehensive description of electromagnetic wave reflection at the interface and absorption within biological tissues across the frequency spectrum.

\subsubsection{Dry Skin}
\begin{figure}[!h]
\vskip 0.0cm
\centering
\includegraphics[width=2.8in]{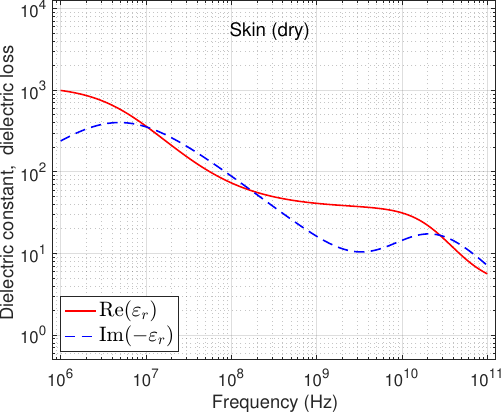}\hspace{0.5cm}
\includegraphics[width=2.8in]{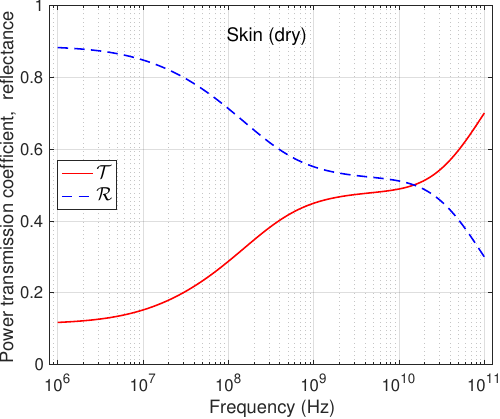} \\[1ex]
\includegraphics[width=2.8in]{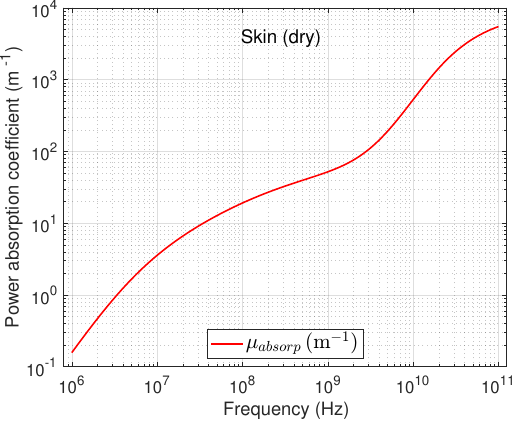}\hspace{0.5cm}
\includegraphics[width=2.8in]{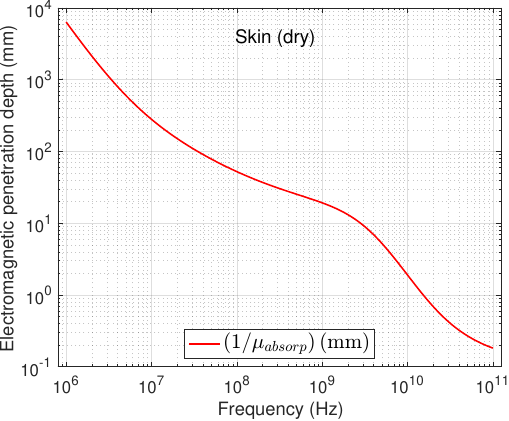}
\vskip -0.0cm
\caption{Dielectric properties of dry skin as functions of frequency (1 MHz--100 GHz). 
Top left: Real and negative imaginary parts of the relative permittivity (dielectric constant and dielectric loss). 
Top right: Power transmission coefficient $\mathcal{T}$ and reflectance $\mathcal{R}$. 
Bottom left: Power absorption coefficient $\mu_\text{absorp}$. Bottom right: Electromagnetic penetration depth ($ 1/\mu_\text{absorp}$).}
\label{fig_1}
\end{figure}

The six dielectric quantities discussed above are presented for dry skin in Figure \ref{fig_1} across the frequency range of 1 MHz to 100 GHz. 

The top-left panel of Figure~\ref{fig_1} illustrates how the frequency-dependent complex permittivity of dry skin governs both interface power coupling and in-tissue attenuation. As the frequency increases from the MHz to the GHz range, the relative permittivity decreases by orders of magnitude, thereby reducing the impedance mismatch
at the air-tissue interface. Consequently, the reflectance $\mathcal{R}$ decreases while the power transmission coefficient $\mathcal{T}$ increases, as shown in the top-right panel of  Figure~\ref{fig_1}. 

This trend indicates that at higher frequencies, a larger fraction of incident power penetrates the tissue. Inside the tissue, however, the power absorption coefficient is given by $\displaystyle 
\mu_\text{absorp}(\omega)=\frac{2\omega}{c_0} \text{Im}(-\sqrt{\varepsilon_r(\omega)})$, which contains a factor of $\omega$. Although $\varepsilon_r(\omega)$ decreases with 
$\omega$, the linear factor of $\omega$ more than compensates for this decline,
rendering $\displaystyle \mu_\text{absorp}(\omega)$ an increasing function of frequency. 
Consequently, the electromagnetic penetration depth, $1/\mu_\text{absorp}$, 
decreases from over 1 m at 1 MHz to below 1 mm at 100 GHz, as illustrated
in the two bottom panels of Figure~\ref{fig_1}. 

Taken together, Figure~\ref{fig_1} highlights a transition from reflection-limited coupling at low frequencies (where most power is reflected at the interface) to absorption-limited penetration at high frequencies (where more power penetrates the interface but is rapidly attenuated). 
%
\subsubsection{Wet Skin}
\begin{figure}[!h]
\vskip 0.1cm
\centering
\includegraphics[width=2.8in]{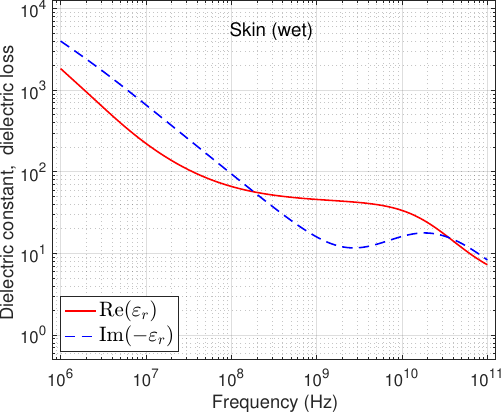}\hspace{0.5cm}
\includegraphics[width=2.8in]{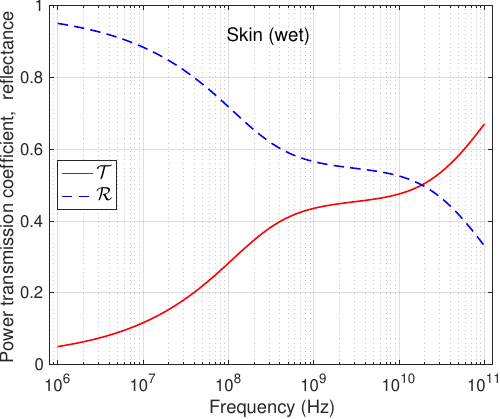} \\[1ex]
\includegraphics[width=2.8in]{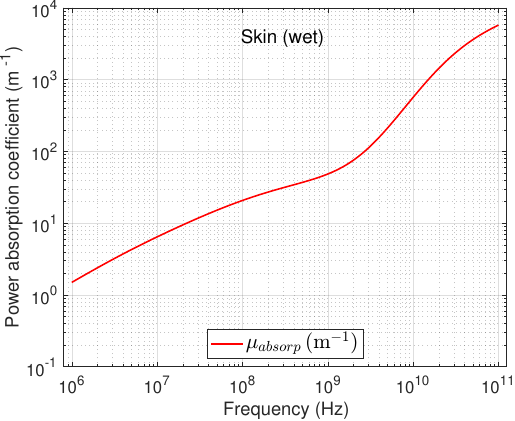}\hspace{0.5cm}
\includegraphics[width=2.8in]{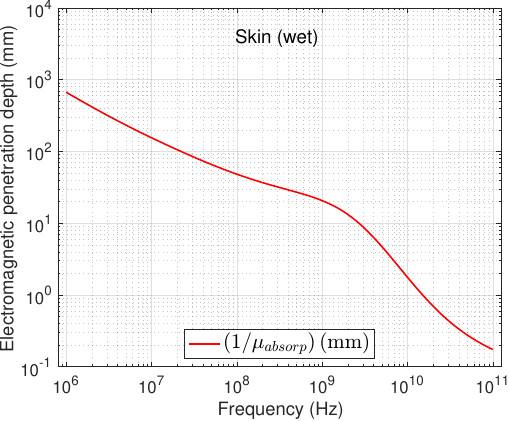}
\vskip -0.0cm
\caption{
Dielectric properties of wet skin as functions of frequency (1 MHz--100 GHz). 
Top left: Real and negative imaginary parts of the relative permittivity (dielectric constant and dielectric loss). 
Top right: Power transmission coefficient $\mathcal{T}$ and reflectance $\mathcal{R}$. 
Bottom left: Power absorption coefficient $\mu_\text{absorp}$. Bottom right: Electromagnetic penetration depth ($ 1/\mu_\text{absorp}$).}
\label{fig_2}
\end{figure}

The frequency-dependent dielectric behavior of wet skin is illustrated in Figure ~\ref{fig_2}, which highlights how hydration amplifies both the dispersive and lossy character of the tissue, shifting the exposure profile compared to the dry-skin case. Across the 1 MHz--100 GHz range, wet skin exhibits higher dielectric loss and generally larger permittivity at low-to-mid frequencies. This increases the impedance mismatch with air, and therefore drives stronger interface reflection at low frequencies where the reflectance $\mathcal{R}$ is near unity and the transmittance $\mathcal{T}$ is correspondingly small with the MHz decade. 

As frequency increases and the relative permittivity decreases, the mismatch is reduced, and the curves trend toward greater coupling into tissue (decreasing $\mathcal{R}$ and increasing $\mathcal{T}$). However, this enhanced coupling is  only half of the story. 
The bottom panels of Figure ~\ref{fig_2} emphasize that, once power is transmitted, wet tissue attenuates it more aggressively. The absorption coefficient $\mu_\text{absorp}$ rises sharply with frequency, and the corresponding penetration depth $1/\mu_\text{absorp}$ is compressed relative to dry skin across the entire band. Figure~\ref{fig_2} demonstrates that hydration tends to reduce surface coupling at low frequencies via higher reflectance, while simultaneously localizing deposited energy closer to the surface at higher frequencies. Consequently, moisture context is a dominant source of variability in both sensing/communication performance and thermal loading predictions.

\subsubsection{Lens Cortex}
\begin{figure}[!h]
\vskip 0.1cm
\centering
\includegraphics[width=2.8in]{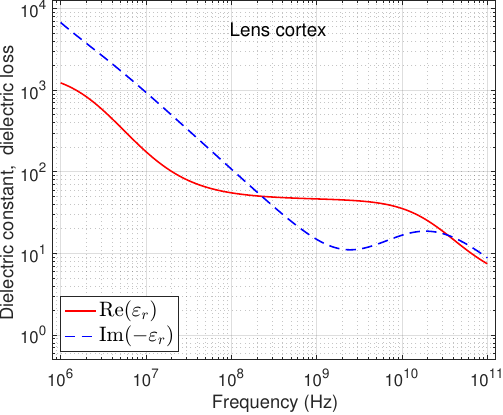}\hspace{0.5cm}
\includegraphics[width=2.8in]{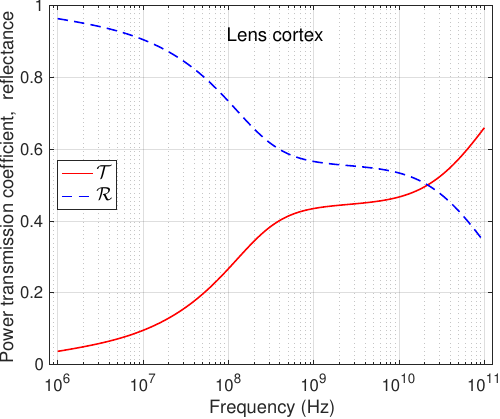} \\[1ex]
\includegraphics[width=2.8in]{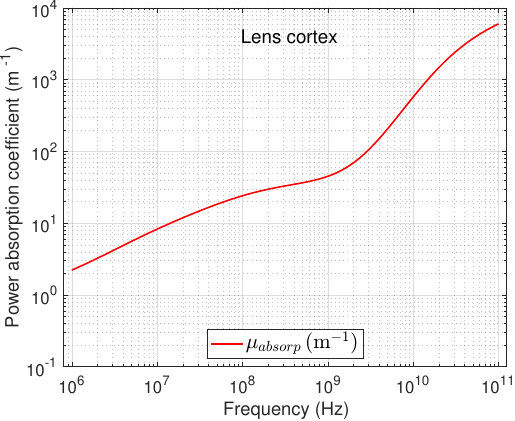}\hspace{0.5cm}
\includegraphics[width=2.8in]{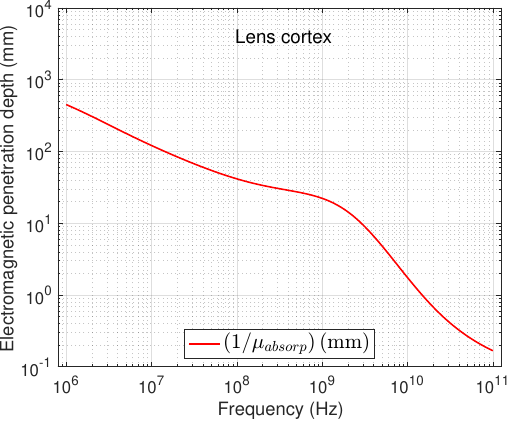}
\vskip -0.0cm
\caption{
Dielectric properties of lens cortex as functions of frequency (1 MHz--100 GHz). 
Top left: Real and negative imaginary parts of the relative permittivity (dielectric constant and dielectric loss). 
Top right: Power transmission coefficient $\mathcal{T}$ and reflectance $\mathcal{R}$. 
Bottom left: Power absorption coefficient $\mu_\text{absorp}$. Bottom right: Electromagnetic penetration depth ($ 1/\mu_\text{absorp}$).}
\label{fig_3}
\end{figure}

The coupling and attenuation properties for the lens cortex are illustrated in Figure~\ref{fig_3}, demonstrating a highly dispersive dielectric response from 1~MHz to 100~GHz. Relative permittivity 
decreases as frequency increases, which alters the wave impedance of the lens tissue. This leads to a severe air-tissue impedance mismatch at low frequencies, resulting in near-unity power reflectance ($\mathcal{R}$) and minimal power transmission ($\mathcal{T}$); conversely, increasing frequency into the GHz range reduces this mismatch, leading to lower $\mathcal{R}$ and higher $\mathcal{T}$, thus indicating greater power coupling into tissue at higher frequencies.

However, Figure~\ref{fig_3} also indicates that increased transmission does not imply deep propagation. The absorption coefficient rises significantly with frequency, causing penetration depth to shrink from tens of centimeters at low frequencies to millimeters and below in the MMW decade. This means that as frequency increases, energy deposition becomes progressively more localized near the entry surface of the lens tissue, with the MMW band producing highly superficial absorption compared to lower RF and microwave frequencies.
%

\subsubsection{Non-infiltrated Fat}
\begin{figure}[!h]
\vskip 0.1cm
\centering
\includegraphics[width=2.8in]{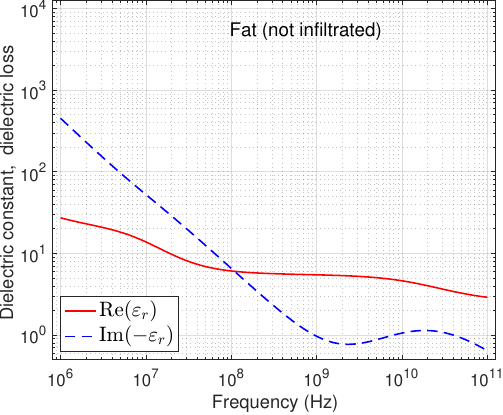}\hspace{0.5cm}
\includegraphics[width=2.8in]{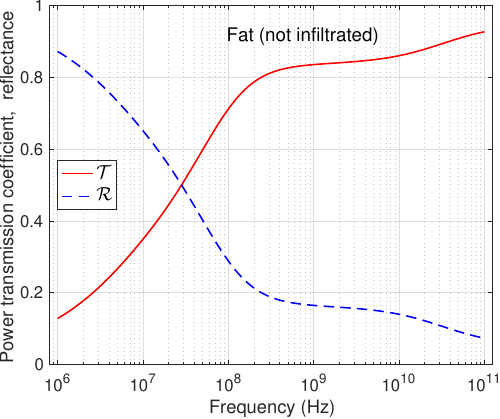} \\[1ex]
\includegraphics[width=2.8in]{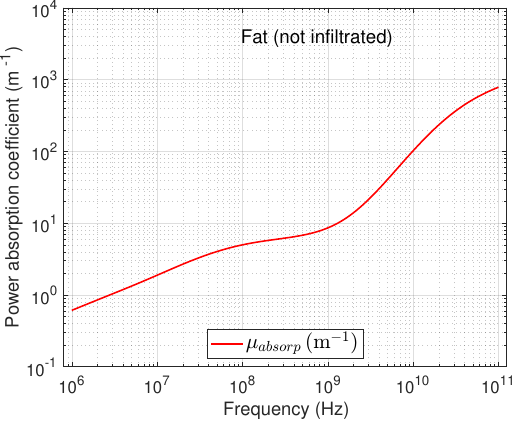}\hspace{0.5cm}
\includegraphics[width=2.8in]{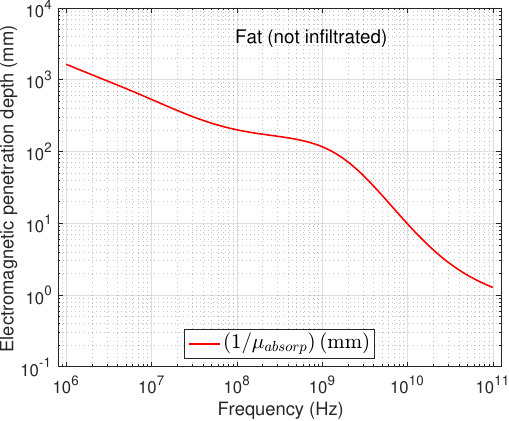}
\vskip -0.0cm
\caption{
Dielectric properties of non-infiltrated fat as functions of frequency (1 MHz--100 GHz). 
Top left: Real and negative imaginary parts of the relative permittivity (dielectric constant and dielectric loss). 
Top right: Power Transmission coefficient $\mathcal{T}$ and reflectance $\mathcal{R}$. 
Bottom left: Power absorption coefficient $\mu_\text{absorp}$. Bottom right: Electromagnetic penetration depth ($ 1/\mu_\text{absorp}$).}
\label{fig_4}
\end{figure}

Figure~\ref{fig_4} characterizes non-infiltrated fat as a low-water-content, relatively low-loss dielectric across the 1~MHz--100~GHz band.  
The real part of the permittivity remains much smaller than that of skin or organ tissues, ranging from the tens at low frequencies to single digits at the higher end, while the loss term declines rapidly into the microwave range. This lower permittivity reduces the impedance contrast with air as frequency increases, a trend reflected in the interface power coefficients: the reflectance 
$\mathcal{R}$ decreases steadily, while the transmittance 
$\mathcal{T}$ rises to dominate through the MMW regime. Consequently, the interface becomes increasingly transmissive with frequency, making power coupling in fat significantly less limited by reflection than in high-permittivity tissues.

At the same time, the bottom panels of Figure~\ref{fig_4} show that penetration is ultimately governed by attenuation: $\mu_\text{absorp}$ increases strongly with frequency, causing the corresponding penetration depth to drop from very large values at low frequencies to the millimeter scale in the MMW range. Compared to highly hydrated and conductive tissues, fat maintains significantly larger penetration depths across most of the spectrum, consistent with its lower dielectric loss. This makes non-filtrated fat an ideal “contrast case” in the tissue set: even when the boundary becomes highly transmissive at high frequencies, the dominant trend remains a transition toward more superficial energy deposition, though this transition occurs less aggressively than in more hydrated, high-loss tissues.

\subsubsection{Liver}
\begin{figure}[!h]
\vskip 0.1cm
\centering
\includegraphics[width=2.8in]{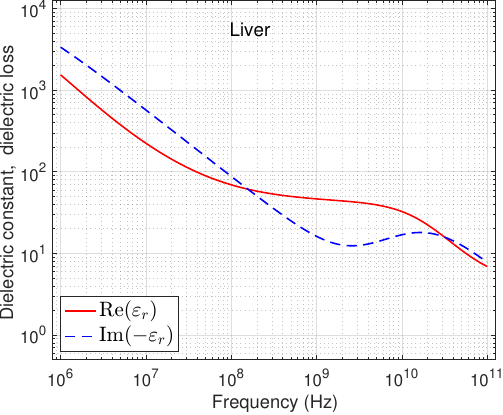}\hspace{0.5cm}
\includegraphics[width=2.8in]{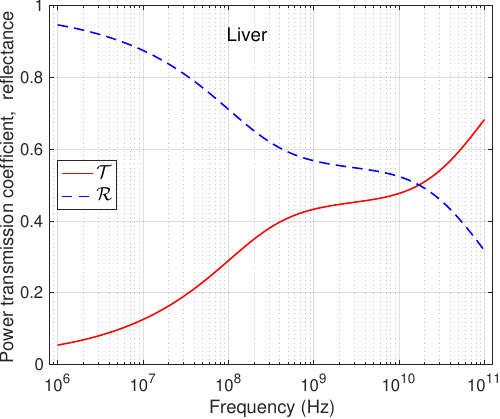} \\[1ex]
\includegraphics[width=2.8in]{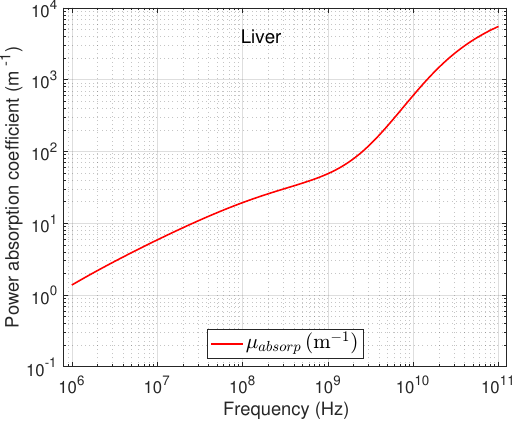}\hspace{0.5cm}
\includegraphics[width=2.8in]{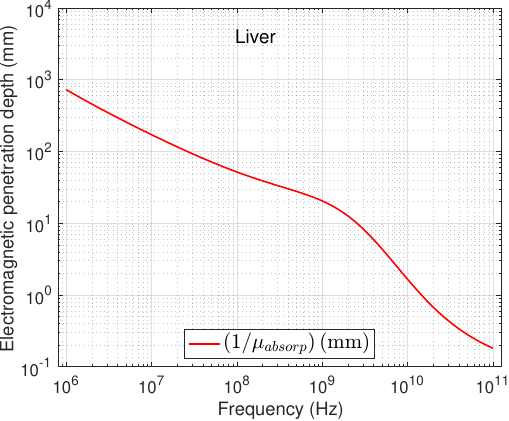}
\vskip -0.0cm
\caption{
Dielectric properties of liver as functions of frequency (1 MHz--100 GHz). 
Top left: Real and negative imaginary parts of the relative permittivity (dielectric constant and dielectric loss). 
Top right: Power transmission coefficient $\mathcal{T}$ and reflectance $\mathcal{R}$. 
Bottom left: Power absorption coefficient $\mu_\text{absorp}$. Bottom right: Electromagnetic penetration depth ($ 1/\mu_\text{absorp}$).}
\label{fig_5}
\end{figure}

Figure~\ref{fig_5} depicts liver acting as a high-permittivity, high-loss tissue, consistent with its high water and ionic content. Both the real part of the permittivity and the loss term are very large at low frequencies and decrease substantially with increasing frequency, indicating strong dispersion across the band. This produces a clear interface signature: at MHz frequencies, the air-liver mismatch is extreme, so the reflectance $\mathcal{R}$ is near unity and the transmittance $\mathcal{T}$ is small. As frequency increases and the permittivity drops, the mismatch relaxes and $\mathcal{T}$ increases as $\mathcal{R}$ falls, meaning a larger fraction of incident power can penetrate the liver in the MMW regime.

However, the liver's key distinguishing feature is its behavior after entry: the absorption coefficient $\mu_\text{absorp}$ grows rapidly with frequency, causing the penetration depth to collapse from approximately 1 meter at 1 MHz to sub-millimeter or millimeter scales in the MMW band. Thus, compared to fat, the liver shifts much more decisively into an absorption-limited regime at high frequencies; although the interface becomes more transmissive, the electromagnetic power is attenuated over a very small depth, resulting in strongly localized near-surface deposition. This contrast between fat and liver is particularly useful for emphasizing the role of tissue water content: it simultaneously dictates boundary coupling (via impedance) and depth localization (via loss), significantly impacting predictions of where power is deposited across the frequency spectrum.
%
\subsubsection{Muscle}
\begin{figure}[!h]
\vskip 0.1cm
\centering
\includegraphics[width=2.8in]{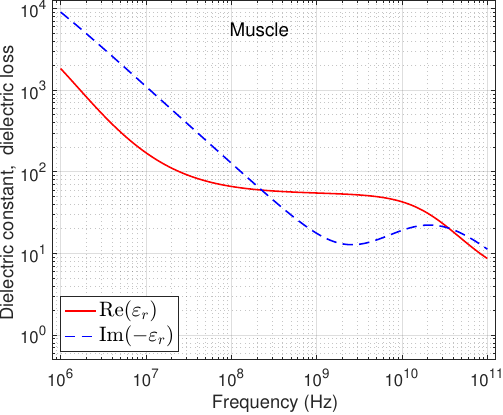}\hspace{0.5cm}
\includegraphics[width=2.8in]{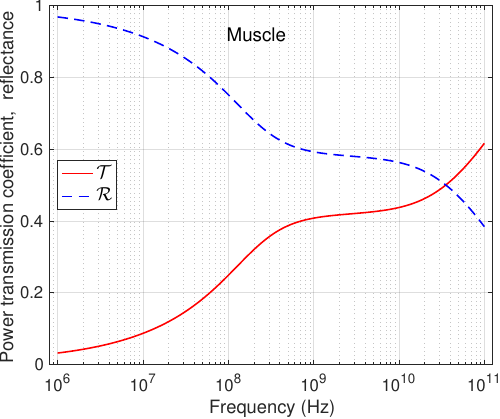} \\[1ex]
\includegraphics[width=2.8in]{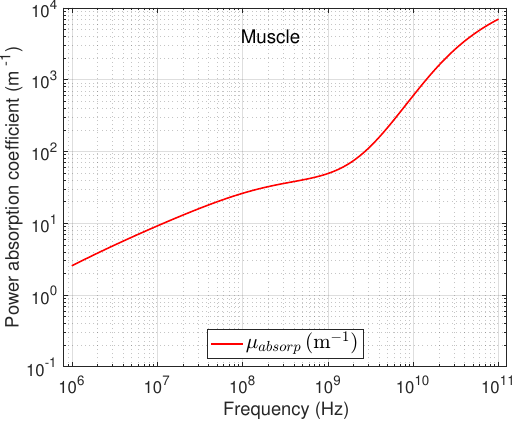}\hspace{0.5cm}
\includegraphics[width=2.8in]{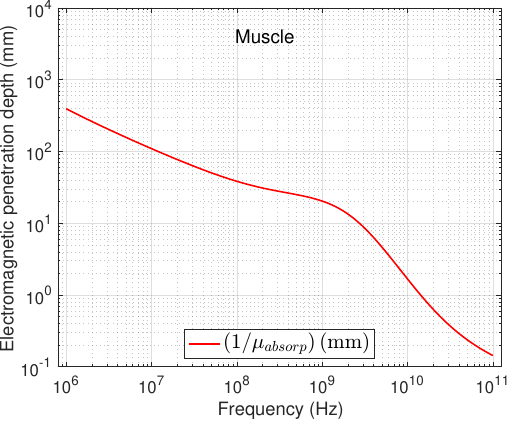}
\vskip -0.0cm
\caption{
Dielectric properties of muscle as functions of frequency (1 MHz--100 GHz). 
Top left: Real and negative imaginary parts of the relative permittivity (dielectric constant and dielectric loss). 
Top right: Power transmission coefficient $\mathcal{T}$ and reflectance $\mathcal{R}$. 
Bottom left: Power absorption coefficient $\mu_\text{absorp}$. Bottom right: Electromagnetic penetration depth ($ 1/\mu_\text{absorp}$).}
\label{fig_6}
\end{figure}

The dielectric properties of muscle tissue across the 1~MHz--100~GHz frequency spectrum are presented in Figure~\ref{fig_6}. 

%
As a high-water-content tissue, muscle exhibits a strongly dispersive response, with both the real permittivity and dielectric loss decreasing significantly as frequency increases. Similar to the liver and skin, the air–muscle impedance mismatch is most severe at low frequencies, resulting in high reflectance ($\mathcal{R} \approx 1$) and minimal power coupling ($\mathcal{T} << 1$) in the MHz range. However, as the relative permittivity drops through the GHz and MMW decades, the interface becomes increasingly transmissive. 

Conversely, the bottom panels of Figure~\ref{fig_6} illustrate that while more power penetrates the interface at higher frequencies, the absorption coefficient 
increases dramatically. This leads to a rapid collapse in penetration depth: from tens of centimeters in the MHz range to tens of millimeters in the low-GHz regime, and finally to millimeter and sub-millimeter scales in the MMW band. This trend indicates that energy deposition becomes highly localized near the surface at high frequencies. 

\subsubsection{Dielectric Properties Across Six Tissues}
\begin{figure}[!h]
\vskip 0.1cm
\centering
\includegraphics[width=2.8in]{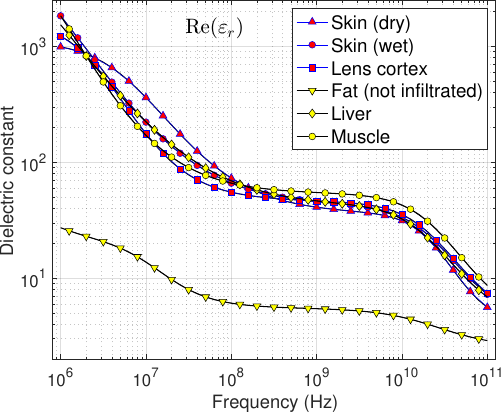} \\[1ex]
\includegraphics[width=2.8in]{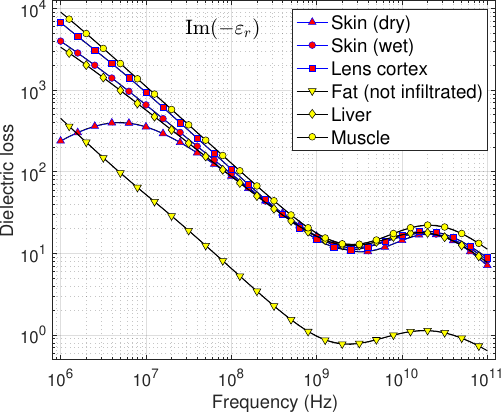}
\vskip -0.0cm
\caption{Frequency-dependent dielectric constants (top) and dielectric losses (bottom) for the six biological tissues.}
\label{fig_7}
\end{figure}

Figure~\ref{fig_7} provides a comparative summary of the dielectric parameters that govern the trends of coupling and penetration across all six biological tissues.  

By presenting these properties on a unified scale, the contrast between high-water-content tissues (such as muscle, liver, and wet skin) and lower-water-content tissues (such as non-infiltrated fat) is clearly highlighted. As shown in the top panel of Figure~\ref{fig_7}, the five more water-rich tissues (dry skin, wet skin, lens cortex, liver, and muscle) largely cluster together across the 1~MHz–-100~GHz range. These tissues exhibit very large $\text{Re}(\varepsilon_r)$ values at low frequencies (on the order of $10^3$) that fall by roughly two orders of magnitude into the GHz decade, indicating strong dielectric dispersion and a substantial frequency-dependent shift in effective wave impedance. In contrast, fat remains a distinct outlier with a substantially lower $\text{Re}(\varepsilon_r)$, ranging from a few tens at MHz to single digits at high frequencies. This implies a much smaller departure from free-space impedance and, consequently, significantly different interface behavior compared to the more hydrated tissues.

The bottom panel of Figure~\ref{fig_7} demonstrates that this grouping persists in the dielectric loss, $\text{Im}(-\varepsilon_r)$. Muscle and lens cortex are among the most lossy, wet skin is consistently lossier than dry skin, and fat remains the lowest-loss tissue across the entire band. All tissues exhibit a decrease in dielectric loss from the MHz range into the mid-GHz range, followed by a broad high-frequency uptick in the tens-of-GHz decade for the water-rich tissues. This behavior is consistent with the high-frequency relaxation processes that become more influential in that regime. 

\begin{figure}[!h]
\vskip 0.1cm
\centering
\includegraphics[width=2.8in]{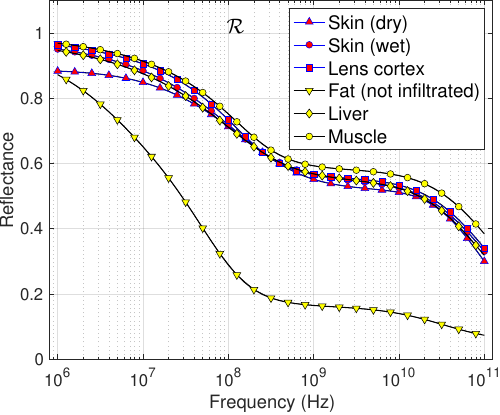} \\[1ex]
\includegraphics[width=2.8in]{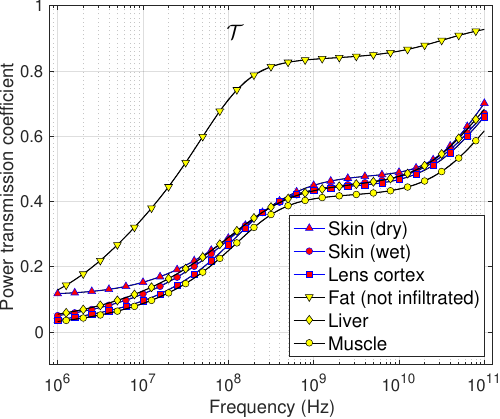}
\vskip -0.0cm
\caption{Comparison of the power reflectance (top) and power transmission coefficient (bottom) for the six biological tissues across the 1~MHz--100~GHz frequency range.}
\label{fig_8}
\end{figure}

Figure \ref{fig_8} presents a comparative analysis of the interface power coupling for all six biological tissues, depicting the frequency-dependent reflectance $\mathcal{R}$ (top) and power transmission coefficient 
$\mathcal{T}$ (bottom). 

A clear trend emerges across all tissues: they become less reflective and more transmissive as frequency increases. This is consistent with the strong dispersion in $\varepsilon_r$ shown earlier: as the relative permittivity 
$\varepsilon_r $ decreases with frequency, the tissue impedance approaches that of free space, thereby reducing the air–tissue mismatch. The five more water-rich tissues (dry and wet skin, lens cortex, liver, and muscle) cluster tightly together, maintaining high reflectance at low frequencies before gradually dropping to a broad mid-band plateau and falling again in the upper tens of GHz. Within this cluster, wet skin tends to be slightly more reflective (and less transmissive) than dry skin, consistent with hydration increasing effective permittivity and loss. Similarly, muscle and lens cortex generally rank among the most reflective tissues due to their higher water and ionic content.

In contrast, fat remains a significant outlier: its reflectance decreases much more rapidly with frequency, while its transmittance rises earlier and remains substantially higher than that of the other tissues across most of the band. This behavior is a direct signature of fat’s much lower permittivity, which produces a smaller impedance contrast with air. Consequently, a larger fraction of incident power couples into the fat at the boundary. Importantly, Figure~\ref{fig_8} examines the power balance solely at the interface. It shows that tissue type and hydration levels significantly alter the amount of power that passes into the medium. These findings underscore why accurate predictions of delivered power and subsequent heating must account for the interplay between boundary coupling and in-tissue attenuation. 

\begin{figure}[!h]
\vskip 0.1cm
\centering
\includegraphics[width=2.8in]{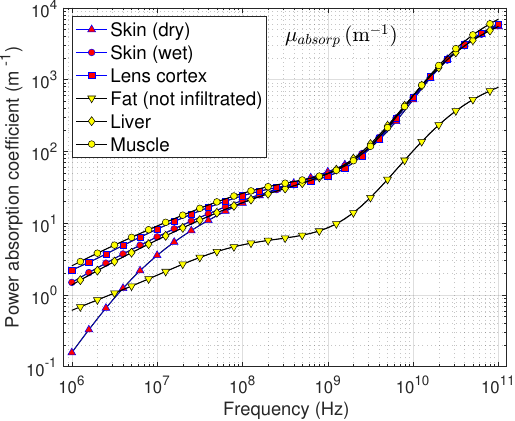} \\[1ex]
\includegraphics[width=2.8in]{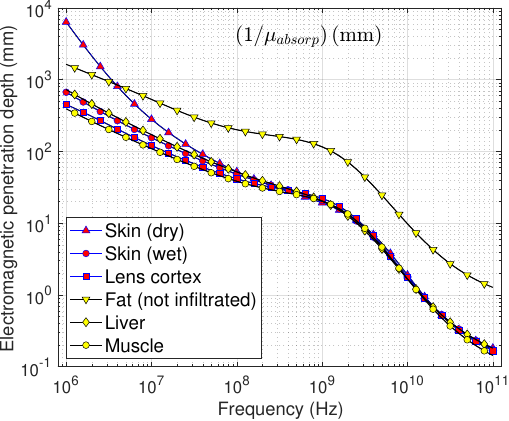}
\vskip -0.0cm
\caption{Comparison of the power absorption coefficient (top) and 
the electromagnetic penetration depth (bottom) for the six biological tissues across the 1~MHz--100~GHz range.}
\label{fig_9}
\end{figure}

%
Figure~\ref{fig_9} synthesizes the tissue-specific results into a unified “attenuation and depth” comparison, illustrating that water and ionic content are the dominant drivers of in-medium loss across the 1~MHz–-100~GHz band. It is important to note that the power absorption coefficient, 
$\displaystyle \mu_\text{absorp}(\omega)=
\frac{2\omega}{c_0} \text{Im}(-\sqrt{\varepsilon_r(\omega)})$,
is different from the dielectric loss, $\text{Im}(-\varepsilon_r(\omega))$. 
At a fixed frequency $\omega$ and dielectric constant $\text{Re}(\varepsilon_r(\omega))$, an increase in dielectric loss leads directly to 
an increase in $\mu_\text{absorp}(\omega)$. However, the overall frequency dependence of $\mu_\text{absorp}(\omega)$ is dominated by the linear factor $\omega$
in the expression, rather than the variations in permittivity alone. 

The top panel of Figure~\ref{fig_9} shows that the power absorption coefficients $\mu_\text{absorp}$ for dry/wet skin, lens cortex, liver, and muscle exhibit tight clustering and increase monotonically with frequency, with a noticeable steepening into the upper-GHz and MMW decades. This indicates that once energy couples into these more hydrated tissues, attenuation becomes progressively stronger and quickly dominates the propagation physics. In contrast, fat remains a low-loss outlier, with $\mu_\text{absorp}$ values substantially below those
of the clustered group.

The bottom panel of Figure~\ref{fig_9} depicts the electromagnetic penetration depth $1/\mu_\text{absorp}$, 
which decreases rapidly with frequency for all tissues, but the same clustering and outlier structure persists. The water-rich tissues fall from macroscopic scales at low frequencies to centimeter, millimeter, and sub-millimeter depths as the frequency approaches the MMW range. This implies that power deposition is increasingly limited to a thin surface layer, even when the interface transmission is favorable (as seen in Figure~\ref{fig_8}). Fat again exhibits systematically larger penetration depths, meaning that at any given frequency, it allows electromagnetic 
power to persist over longer distances than the other tissues, though it also transitions to millimeter-scale depths at MMW frequencies. Overall, Figure~\ref{fig_9} provides a clear energy-localization message: increasing the frequency shifts deposition toward the surface in every tissue, though fat delays this transition relative to more aqueous tissues.

\section{Concluding Remarks}

This work developed an analytical plane-wave framework based on the one-dimensional (1D) Maxwell equation for interpreting electromagnetic reflection and transmission at the air-tissue interface, as well as attenuation within the tissue. 
The electromagnetic fields were solved across a broad frequency span 
(1~MHz–-100~GHz) using the frequency-dependent relative permittivity derived from the Cole-Cole dielectric model. Specifically, these interactions were examined for six biological tissues using parameter values reported by Gabriel \emph{et al.} \cite{Gabriel_1996b}. 

We partition the electromagnetic interactions into two distinct processes: 
\begin{enumerate}
\item Interface power coupling, governed by
impedance mismatch (quantified by power reflectance $\mathcal{R}$ and transmittance $\mathcal{T}$).
\item In-medium attenuation, characterized by the power absorption coefficient $\mu_\text{absorp}$ and penetration depth $1/\mu_\text{absorp}$.
\end{enumerate}

This framework provides an intuitive “power budget” perspective, revealing a systematic trend as frequency increases. While dielectric dispersion reduces the impedance mismatch and generally increases transmitted power at the air–tissue boundary, attenuation strengthens rapidly and drives penetration depths toward the millimeter and sub-millimeter scales in the MMW band, effectively localizing deposited power within a thin surface layer. 

Across tissue types studied, the primary source of variability is the complex relative permittivity signature associated with water and ionic content. Highly hydrated tissues, including skin, liver, muscle, and lens cortex,  cluster in their reflectance, transmittance, and attenuation trends, exhibiting high reflectance at low frequencies and intense attenuation at high frequencies. In contrast, fat remains a distinct outlier characterized by substantially higher boundary transmission, a lower attenuation rate, and deeper penetration across the 10~MHz to 100~GHz range. Hydration further modulates both components of the power budget, with wet skin showing greater loss and shallower penetration than dry skin. 

While the present analysis assumes a homogeneous half-space and normal incidence, it establishes clear first-order expectations for how frequency and tissue composition jointly shape coupling and depth localization. Future extensions should incorporate multilayered anatomy, oblique incidence, and polarization effects, as well as thermophysical transport, to translate these electromagnetic length scales into predictive heating and exposure metrics for realistic geometries and applications.
\clearpage
\noindent{\bf \large Acknowledgment and disclaimer}

\noindent \indent
The authors gratefully acknowledge the support of the Joint Intermediate Force Capabilities Office of the U.S. Department of Defense and the Naval Postgraduate School. The views and conclusions expressed in this document are those of the authors and do not reflect the official policy or position of the Department of Defense or the U.S. Government.

\end{document}